\documentclass[twocolumn,aps,showpacs]{revtex4-1}
\usepackage{amsfonts}
\usepackage{amsmath}
\usepackage{amssymb}
\usepackage{txfonts}
\usepackage{pxfonts}
\usepackage{graphicx,bm,units,yfonts}
\usepackage{subfigure}
\usepackage[table]{xcolor}
\usepackage{hyperref}

\newcommand{\RM}{{\mathbb R}}

\newcommand{\ZM}{{\mathbb Z}}

\newcommand{\Tt}{{\mathcal T}}

\newcommand{\Pp}{{\mathcal P}}

\newcommand{\Ll}{{\mathcal L}}

\begin{document}

\title{Topological Gaps by Twisting}

\author{Matheus Rosa, Massimo Ruzzene}

\affiliation{Department of Mechanical Engineering, University of Colorado Boulder, Boulder, CO, USA}

\author{Emil Prodan}
\affiliation{Department of Physics, Yeshiva University, New York, NY, USA}

\begin{abstract}
It is shown that twisted $n$-layers have an intrinsic degree of freedom living on $2n$-tori, which is the phason supplied by the relative slidings of the layers and that the twist generates pseudo magnetic fields. As a result, twisted $n$-layers host intrinsic higher dimensional topological phases and those characterized by  second Chern numbers can be found in a twisted bi-layer. Indeed, our investigation of phononic lattices with interactions modulated by a second twisted lattice reveals Hofstadter-like spectral butterflies in terms of the twist angle, whose gaps carry the predicted topological invariants. Our work demonstrates how multi-layered systems are virtual laboratories for studying the physics of higher dimensional quantum Hall effect and how to generate topological edge chiral modes by simply sliding the layers relative to each other. In the context of classical metamaterials, both photonic and phononic, these findings open a path to engineering topological pumping via simple twisting and sliding.
\end{abstract}


\maketitle

\section{Introduction} 

Engineering topological  states using aperiodic principles is an extremely active area of research, spread over different fields such as condensed matter~\cite{YoshidaPRB2013,HeEPL2015,ProdanPRB2015,TranPRB2015,VidalPRB2016,FulgaPRL2016,
CollinsNature2017,AgarwalaPRL2017,HuangPRL2018,BournJPA2018,
VarjasPRL2019,DevakulPRB2019,PaiPRB2019,KellendonkAHP2019,
ChenArxiv2019,IliasovPRB2020,FremlingPRB2020,
HuangPRB2020,ChenPRL2020,DuncanPRB2020}, photonics~\cite{KrausPRL2012,KrausPRL2013,VerbinPRL2013,Vardeny2013,
TanesePRL2014,VerbinPRB2015,HuPRX2015,BandresPRX2016,
DareauPRL2017,BabouxPRB2017,ZilberbergNature2018,
KollarNature2019,KollarCMP2020,CarusottoNatPhys2020,
SchultheissAPX2020, YangLSA2020,ZhouLSA2020}, acoustics~\cite{ApigoPRL2019,NiCP2019,ChengPre2020} and mechanics~\cite{MitchellNature2018,MartinezPTRSA2018,ApigoPRM2018,RosaPRL2019,PalNJP2019, ZhouPRX2019,XiaPRAppl2020,RivaPRB2020,RivaPRB2020b,XiaArxiv2020}.  Common to all these is the existence of an intrinsic degree of freedom, the phason, which in many instances is experimentally accessible and fully controllable. The phason space augments the physical space and supplies additional virtual dimensions \cite{ProdanPRB2015}, hence enabling physical phenomena beyond what can be ordinarily observed in our physical space. In particular, the phason can be used as an adiabatic knob to engineer topological pumping, which is one of the main applications of the formalism, as evidence by the literature cited above.

Twisted graphene bilayers were recently observed to have exceptional spectral characteristics \cite{MorellPRB2010,BistritzerPNAS2011,CarrPRB2017} that host extremely rich single- and many-body physics \cite{Cao1Nature2018,Cao2Nature2018,SeylerNature2019,TranNature2019}. Exciting phenomena occurring at magic angles have been observed, such as superconductivity in the flat bands~\cite{Cao1Nature2018} of twisted graphene bilayers and, very recently, hyperbolic and elliptic dispersion of polaritions in twisted photonic systems~\cite{HuNature2020}. Definitely, the field of twistronics, consisting of spectral and dynamical engineering by twisting layered materials and meta-materials, is one of the most active research fields at the present time. Interestingly, in \cite{SongPRL2019} it was found that the spectral gaps stabilized by the magic angles carry non-trivial (fragile) topological indices. Another exciting topological finding  is that, under a modest magnetic field, the bilayered graphene opens topological gaps that host correlated Chern insulating states \cite{NuckollsArxiv2020}. 

Moir\'e patterns, as any other aperiodic pattern, can be analyzed, at least formally, with the operator algebraic program \cite{Bellissard1986,Bellissard1995,KellendonkRMP95} pioneered by Bellissard and others, long before the bilayers came to the full attention of the physics community. The power of this formalism was already demonstrated for incommensurate multi-layered systems in \cite{CancesJMP2017}, where a solution of the extremely difficult problem of computing the transport coefficients was given. However, the topological part of this comprehensive program, as applied to Moir\'e patterns, was missing. Zak \cite{ZakPRL1989} and Avron {\it et al} \cite{AvronPRL1983} thought us to look for winding and Chern numbers if an adiabatic parameter is given and lives on circles or tori, but one should contemplate that there is no obvious circle or torus associated with twisted bilayers. The work of Bellissard \cite{Bellissard1986} indicates that such smooth manifolds could come in the form of the hull of the pattern, a concept that he introduced and referenced here as the phason space. For the bilayered systems described below, we compute this hull (more precise, the transversal of the pattern) and find it to be the 2-torus. Following further Bellissard's program, we prove the following statement: if one arranges and couples identical resonator in a periodic lattice and modulates the couplings by a second twisted lattice, the dynamical matrices always land in the non-commutative 4-torus, regardless of the details. This statement completely classifies the dynamics for this class of systems. As an automatic followup, we can announce that the bilayered systems support topological phases characterized by the $2^{\rm nd}$-Chern number, without any fine tunning or external magnetic fields. Furthermore, we show that topological edge chiral modes can be generated by simply sliding the layers relative to each other, hence supplying a robust and effective mechanism for topological pumping. Our analysis can be straightforwardly generalized to multi-layered systems such as tri-layers, where even higher virtual dimensional topological phases appear and the relative slidings of the layers can be achieved in more interesting ways.

\begin{figure*}
\includegraphics[width=\linewidth]{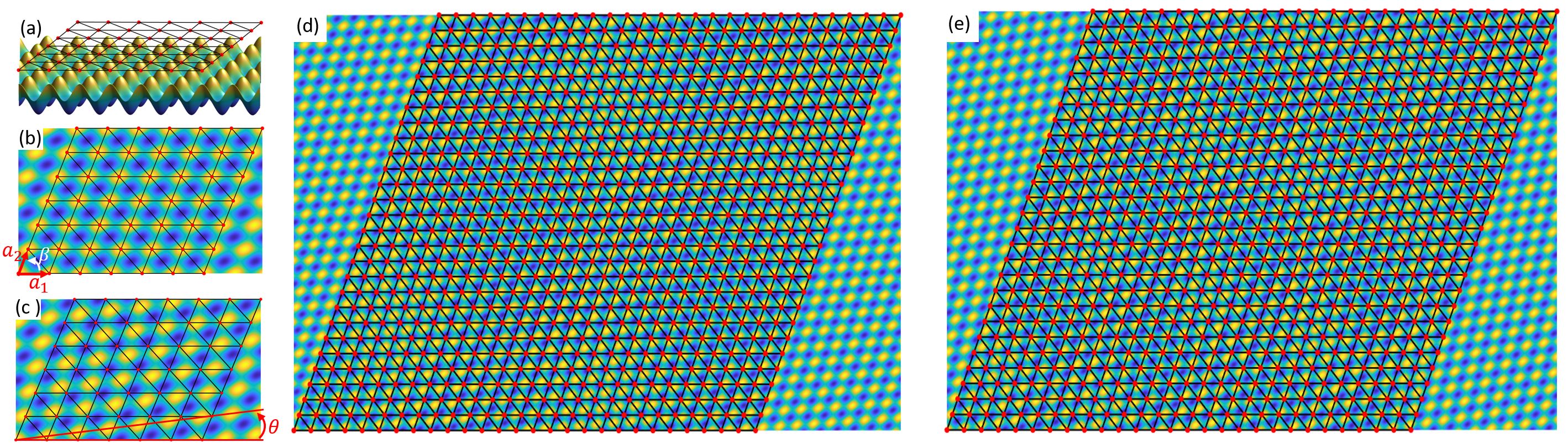}
\caption{\small {\bf Mechanical systems defined.} (a) Mechanical lattice (top layer) with an underlying potential surface (bottom layer). (b) Top view displaying lattice vectors $\bm{a_1}$ and $\bm{a_2}$ separated by an angle $\beta$. In this un-twisted configuration, the lattice sites are aligned with the peaks of the potential. (c) Also in a top view, the potential is twisted by an angle $\theta$ relative to the lattice. (d,e) Examples of truly aperiodic twisted bilayers for $\theta=\pi/50$ and $\theta=\pi/25$, respectively. The figures were generated for $a_2=\sqrt[4]{2} a_1$ and $\beta=\pi/\sqrt{7}$.}
\label{Fig:System}
\end{figure*}

To demonstrate the above principles, we consider three generic twisted phononic systems of different lattice symmetries, for which we map the resonant spectra as a function of twist angle. Without any tuning, the computations reveal Hofstadter-like spectral butterflies and, as we shall see, the integrated density of states (IDS) evaluated inside the spectral gaps gives rise to well defined but intricate patterns of curves. The qualitative shape of these curves change from one example to another, yet we put forward one unifying prediction which fits every single pattern seen in our IDS plots. This prediction follows from the K-theory of the non-commutative 4-torus and the agreement with the numerical computations leave no doubt that the dynamical matrices for these systems fall into that algebra, which is the main statement of our work.

There are important practical consequences of our findings. The phason can be moved on its rightful space by sliding the bilayers relative to each other. The twist angle dictates how many chiral modes are generated by a cycle. This is by far the simplest and easiest way to manipulate the phason of an aperiodic pattern and produce topological pumping (but see also \cite{ChengPre2020}). In the presence of an edge, looping the phason around the fundamental loops of this torus results in topological chiral edge modes. The reader will find in this paper a bulk-boundary principle, tested and confirmed by the numerical observations, which predicts the number of these chiral bands from the values of the bulk topological invariants. By that, we demonstrated that twistronics supplies new ways to generate and manipulate topological edge excitations with unprecedented control and precision. For bilayer graphene, for example, a simple vibration of the layers relative to each other should reveal the existence of the predicted topological edge modes.

On the computational side, let us recall that twisted bilayers away from the special angles are notoriously difficult to deal with because of lack of periodic approximants~\cite{CancesJMP2017}. For the present context, things are made more difficult by the topological edge states which contaminate the bulk spectral gaps. This is a general problem for aperiodic topological systems and is also encountered, for example, in topological quasicrystals \cite{LoringJMP2019,DuncanPRB2020}. We found that the periodic boundary conditions eliminate these topological edge states but impurity-like edge states still persist. The latter, however, have a low density and, as a consequence, the maps of the density of states, as opposed to the spectrum itself, supply remarkably clean pictures of the spectral butterflies \cite{Footnote1}. The quality of the numerical simulations is reflected in sharpness of the integrated density of states inside the spectral gaps, which is the essential numerical outcome that is used  to compare with the theoretical predictions and extract the topological invariants.

\begin{figure*}
\centering
\includegraphics[width=\linewidth]{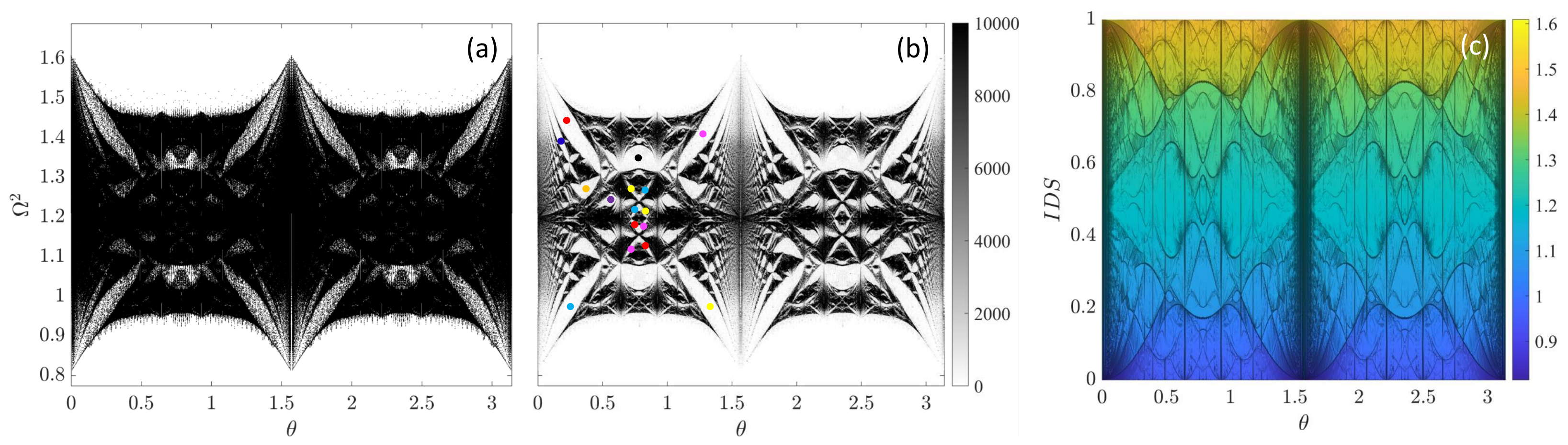}
\caption{{\bf Spectral analysis of the first mechanical system.} \small $\theta$-dependence of (a) resonant spectrum, (b) density of states and (c) integrated density of states, for a square lattice of resonators ($\beta=\pi/2$, $a_1=a_2=1$). The color bar in (c) represents the energy. The larger spectral gaps in panel (b) are identified and labeled for reference.}
\label{Fig:SpecIDS1}
\end{figure*}

\section{Results}

\subsection{The mechanical system defined}

We consider a lattice $\Ll_1$ of identical masses of fixed $(x,y)$ coordinates $\bm r_{\bm n} = n_1 \, \bm a_1 + n_2 \, \bm a_2$, $\bm n=(n_1,n_2) \in \ZM^2$, where $\bm{a_1}$ and $\bm{a_2}$ are arbitrary lattice vectors separated by an angle $\beta$ (Fig.~\ref{Fig:System}). We denote their magnitudes by $a_1$ and $a_2$, respectively. The masses move along the $z$ direction and interact via two-body potentials, while each mass experiences an external potential, represented by the colored surface in Fig.~\ref{Fig:System}. Generically, such system is described by a Lagrangian
\begin{align}
\Ll = &  \sum_{\bm n} \left ( \tfrac{1}{2}m \, \dot z^2_{\bm n} - V_\theta(\bm r_{\bm n}, z_{\bm n} )\right )  \\ \nonumber 
 & \qquad \quad - \tfrac{1}{2} \sum_{\bm n,\bm n'}  W (\bm r_{\bm n} - \bm r_{\bm n'},z_{\bm n}-z_{\bm n'}).
\end{align} 
The functional form of $V_\theta$ is 
\begin{equation}
V_\theta (\bm r,z)=V_0\big (\hat R_\theta [\bm r], z\big ),
\end{equation} 
where $\hat R_\theta$ is the rotation matrix by $\theta$ of the $(x,y)$-plane and $V_0$ is a periodic potential $V_0(\bm r + \bm r_{\bm n},z) = V_0(\bm r,z)$ for all $\bm r_n \in \Ll_1$. We define $\Ll_2 = \hat R_\theta^{-1}[\Ll_1]$ to be the twisted lattice, such that $V_\theta(\bm r + \bm r'_{\bm n}) = V_\theta(\bm r)$ for all $\bm r'_{\bm n} \in \Ll_2$.

In Figs.~\ref{Fig:System}(d,e), we illustrate two configurations of the system, corresponding to a generic lattice $\Ll_1$ and two twist angles $\theta$ such that the periodicity of the system cannot be restored no matter what super-cell is used. The latter can only happen for a discrete set of $\theta$-s, hence, the generic cases are those of purely aperiodic configurations. Our analysis will cover both the special and generic cases on equal footing. Let us recall that a fine sampling of the twists by commensurate angles requires notoriously large supercells \cite{LopesPRB2012} and we want to assure the reader, and especially the experimentalists, that our analysis and conclusions do not rely on any such periodic approximants.    Let us mention that $z$ can be replaced with any other local degree of freedom, such as a rotation angle. In that case, laboratory models of the system introduced above can be implemented, for example, with the systems of magnetically coupled spinners introduced in \cite{ApigoPRM2018,QianPRB2018}. This task, however, is left to the future for now.

In the regime of small oscillations, the dispersion equation of the collective resonant modes takes the form
\begin{align}\label{Eq;EqMotion}
m \Omega^2 \zeta_{\bm n}= & \big (K_{\bm n} + V_\theta''\big (\bm r_{\bm n},\bar z_{\bm n}\big ) \big ) \xi_{\bm n} \\ \nonumber 
& \qquad - W''\big (\bm r_{\bm n} - \bm r_{\bm n'},\bar z_{\bm n}-\bar z_{\bm n'}\big) \, \zeta_{\bm n'}.
\end{align}
Here, $\zeta_{\bm n}=z_{\bm n} - \bar z_{\bm n}$ with $\bar z_{\bm n}$ being the equilibrium $z$-coordinates of the masses and
\begin{equation}
K_{\bm n} = \sum_{\bm n'} W''\big (\bm r_{\bm n} - \bm r_{\bm n'},\bar z_{\bm n}-\bar z_{\bm n'}\big).
\end{equation}
Note that, in general, both the potential and the coupling constants are perturbed by the $\Ll_2$ lattice. This will be considered in our theoretical analysis but left aside in our numerical experiments.

\subsection{Numerical simulations}

In our numerical applications, we considered a short-range two-body interaction such that 
\begin{equation}
W''(\bm r_{\bm n} - \bm r_{\bm n'}) = e^{-3|\bm r_{\bm n} - \bm r_{\bm n'}|^2}.
\end{equation} 
Let us be clear that the pair interactions were not truncated to first nearest neighbors and, instead, all pairs of masses interact even though the interaction might be exponentially small. We chose a potential such that
\begin{equation}
V''_\theta(\bm r_{\bm n}) = 0.1\, \big (\cos\big (\bm b_1 \cdot \hat R_\theta [\bm r_{\bm n}]\big) + \cos\big(\bm b_2 \cdot \hat R_\theta [\bm r_{\bm n}]\big)\big),
\end{equation} 
where $\bm b_i$-s are $\Ll_1$'s reciprocal vectors. The resulting dynamical matrix for the system of equations \eqref{Eq;EqMotion} was exactly diagonalized on a $100 \times 100$ resonator lattice with periodic boundary conditions (PBC), while sampling $\theta$ over 1000 equally spaced points in the interval $[0,\pi]$. The mass $m$ was set to 1.

\begin{figure*}
\centering
\includegraphics[width=\linewidth]{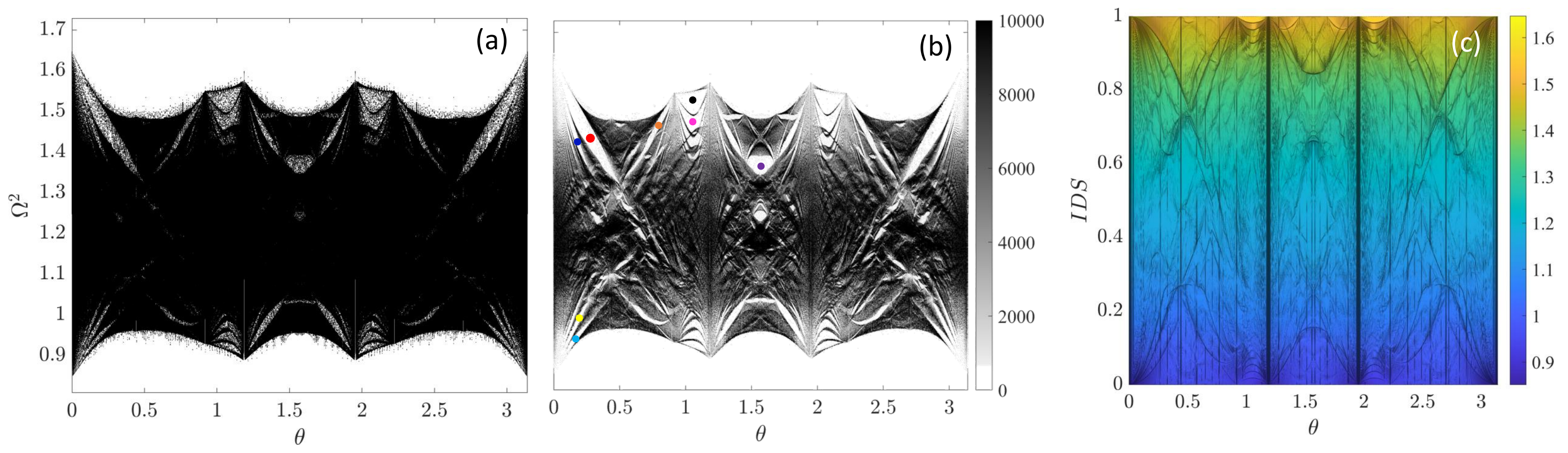}
\caption{\small {\bf Spectral analysis of the second mechanical system.} Same as Fig.~\ref{Fig:SpecIDS1} for the lattice $\beta=\pi/\sqrt{7}$ and $a_1=a_2=1$.}
\label{Fig:SpecIDS2}
\end{figure*}

\begin{figure*}
\centering
\includegraphics[width=\linewidth]{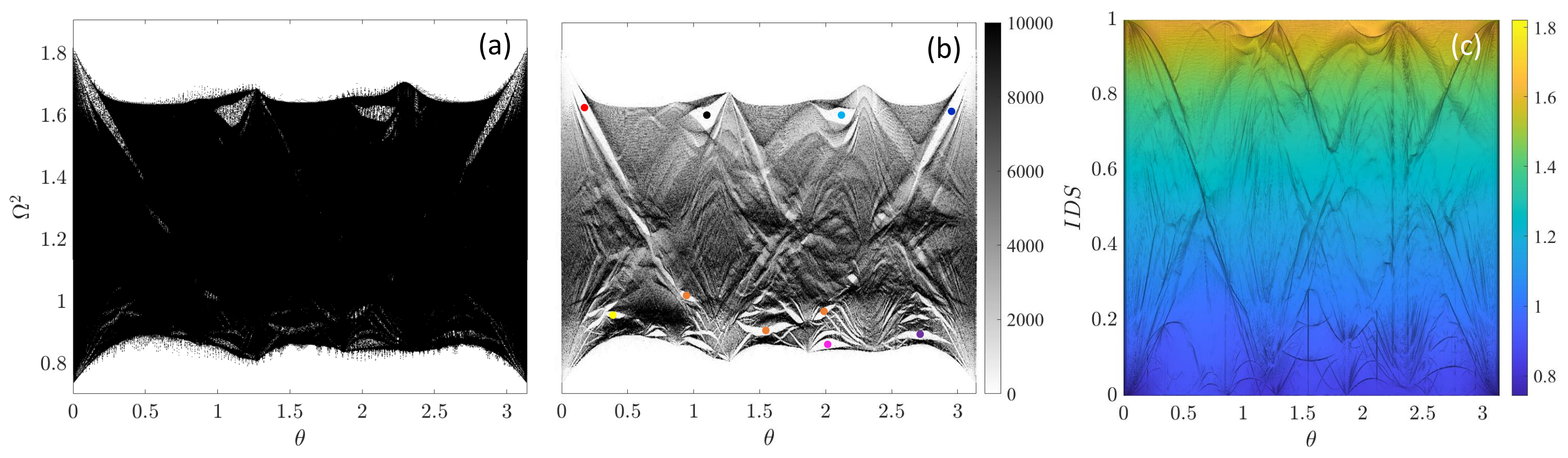}
\caption{\small {\bf Spectral analysis of the third mechanical system.} Same as Fig.~\ref{Fig:SpecIDS1} for the lattice $\beta=\pi/\sqrt{7}$, $a_1=1/\sqrt[4]{2}$ and $a_2=1$.}
\label{Fig:SpecIDS3}
\end{figure*}

We chose three representative lattices such that, at one end, we have a square lattice with a large point group symmetry and, on the other end, a lattice where both $\beta/2\pi$ and $a_1/a_2$ are irrational numbers such that the point group is trivial. The main reason for this is to convince the reader that our statements are independent of the point symmetry of the lattice, as it should for topological phases from class A. A deeper reason for these choices will be revealed in section~\ref{Sec:ThVsNum}. Figs.~\ref{Fig:SpecIDS1}(a),~\ref{Fig:SpecIDS2}(a) and ~\ref{Fig:SpecIDS3}(c) report the resonant spectra of these systems as functions of $\theta$. Large bulk spectral gaps contaminated by edge spectrum are visible in all cases and, overall, the spectra project the same kind of fractality seen in the Hofstadter butterfly \cite{Hofstadter1976}. Let us specify that PBC prevents the topological edge modes but impurity edge states still persist because periodicity is broken by the twisted potential. Nevertheless, PBC are preferred because, while the impurity states still contaminate the bulk gaps, they do not display any spectral flow with $\theta$, as it is evident in all our results.

An important numerical finding is that the spectra can be almost entirely cleared of the edge states contamination by computing the corresponding density of states. This is exemplified in Figs.~\ref{Fig:SpecIDS1}(b),~\ref{Fig:SpecIDS2}(b) and ~\ref{Fig:SpecIDS3}(b). The symmetries of the spectral butterflies are now more apparent. For example, for the square lattice we have reflection symmetries about mid horizontal and vertical lines, as well as $\theta \rightarrow \pi/2 - \theta$. This is why we only labeled gaps in the left half of the spectral butterfly. For the lattice with $a_1=a_2$ and $\beta=\pi/\sqrt{7}$, the reflection symmetry w.r.t. the mid vertical line is still present, and we still focus on spectral gaps from the left side of the spectral butterfly. For the most generic lattice $a_1 \neq a_2$ and $\beta=\pi/\sqrt{7}$, all the symmetries are lifted and we will investigate spectral gaps from all parts of the spectral butterfly.

The most interesting outcomes of our simulations are the integrated density states (IDS), defined as
\begin{equation}
{\rm IDS}(\Omega) = \left . \frac{\big |{\rm Spec}(D) \cap (-\infty, \Omega^2]\big |}{|\Ll_1|} \right |_{\Ll_1 \rightarrow \ZM^2},
\end{equation}
where ${\rm Spec}(D)$ is the spectrum of the dynamical matrix, {\it i.e.} the set of eigenvalues counted with their degeneracies. Throughout, $|\cdot |$ represents the cardinal of a set. The maps of the IDS as function of $\theta$ and $\Omega^2$ are reported in Figs.~\ref{Fig:SpecIDS1}(c),~\ref{Fig:SpecIDS2}(c) and ~\ref{Fig:SpecIDS3}(c) for the three lattices considered in our study. Since the ${\rm IDS}(E)$ is constant when $E$ takes values in the spectral gaps, the 3-dimensional IDS plots have an abrupt variation with respect to $E$ whenever $E$ traverses a gap. In the color maps shown in our figures, these variations appear as abrupt changes of the color and these features were further enhanced by using appropriate lightning. As a result, the values of the IDS inside the gaps can be easily identified by the dark lines visible in all our plots. As one can see, there are drastic changes in the pattern of these lines from one lattice to another. Yet, as we shall see, {\it all} IDS curves in these three figures are described by one unifying equation, where the topological invariants appear as integer coefficients.

\subsection{Theoretical Interpretation}

In this section we explain the features seen in the numerical experiments using the K-theoretic tools developed in \cite{Bellissard1986,Bellissard1995,KellendonkRMP95}. These works demonstrated the existence of a standard formalism, but the statements are not entirely constructive, hence, every new application poses certain computational challenges. As such, it is remarkable that Bellissard's program can be carried out entirely and explicitly for the present context.

{\it Algebra of dynamical matrices.} We encode the degrees of freedom in the vector $|Z\rangle = \sum_{\bm n \in \ZM^2} \zeta_{\bm n} \, |\bm n\rangle$ and transform the dispersion equations into $\omega^2|Z \rangle = D |Z\rangle$ with the dynamical matrix
\begin{equation}
D = \sum_{\bm m,\bm n} w_{\bm m,\bm n}(\Pp) \, |\bm m \rangle \langle \bm n|
\end{equation}
written here in the most generic form. As we shall see, the details of the coupling coefficients are not important for this analysis. What is important is that they are fully determined by the pattern $\Pp$, which consists of the union of the resonator lattice $\Ll_1$ and potential lattice $\Ll_2$. In other words, if the potential and the type of resonators are fixed and no external  intervention is allowed, there are pre-defined functions $w_{\bm m,\bm n}(\Pp)$ of variable $\Pp$ that supply the couplings. Upgrading the coupling coefficients to coupling functions is a strategic point in the theory of dynamics over patterns. For our explicit model, these functions are already specified in \eqref{Eq;EqMotion}. In typical meta-material experiments, these functions can be mapped entirely by exploring the pattern space as it was done, for example, in \cite{ApigoPRM2018}.  Once these functions are cataloged, one can evaluate them on a particular pattern and generate the dynamical matrix. 

The next important observation is Galilean invariance, which says that if we rigidly translate $\Pp$, hence both lattices, the coupling functions  must display the following covariance relations \cite{ProdanJGP2019}:
\begin{equation}
w_{\bm m-\bm a,\bm n- \bm a}(\tau_{\bm a} \Pp) = w_{\bm m,\bm n}(\Pp)  \ \ {\rm or} \  \  w_{\bm m,\bm n}(\Pp) = w_{\bm m- \bm n,\bm 0}(\tau_{\bm n}\Pp),
\end{equation}
where $\tau_{\bm a}\Pp$ is the rigid shift of the pattern which brings the resonator labeled by $\bm a \in \ZM^2$ to the origin. $\tau_{\bm a}\Pp$ is also the pattern seen by an observer which jumped from the origin to the site $\bm a$ of the resonator lattice $\Ll_1$.

After these observations are in place, something magic happens \cite{ProdanJGP2019}. Indeed, we can drop one redundant index and, using $\bm q = \bm m- \bm n$ as well as the shift operator $S_{\bm q} |\bm n \rangle = | \bm n + \bm q \rangle$,  $D$ takes a very particular form
\begin{equation}\label{Eq:Ham2}
D = \sum_{\bm q} S_{\bm q} \sum_{\bm n} w_{\bm q} (\tau_{\bm n} \Pp) \, |\bm n \rangle \langle \bm n |.
\end{equation}  
The extraordinary conclusion is that any Galilean invariant dynamical matrix over $\Pp$ is generated from a small algebra generated by the elementary shift operators $S_i$, $i=1,2$, and by diagonal operators $T_f = \sum_{\bm n} f (\tau_{\bm n} \Pp) \, |\bm n \rangle \langle \bm n |$ with $f$ a function on the space of patterns. Furthermore, one can check the commutation relation (CR)
\begin{equation} 
\sum_{\bm n} f (\tau_{\bm n} \Pp) |\bm n \rangle \langle \bm n | S_{\bm q} = S_{\bm q} \sum_{\bm n} f (\tau_{\bm n+\bm q} \Pp) \, |\bm n \rangle \langle \bm n |,
\end{equation} 
which can be written more compactly as 
\begin{equation}\label{Eq:CommRel}
T_f \, S_{\bm q}  = S_{\bm q} \, T_{f\circ \tau_{\bm q}} \ \ \forall \ \bm q \in \ZM^2.
\end{equation} 

\begin{figure}[b]
\includegraphics[width=0.8\linewidth]{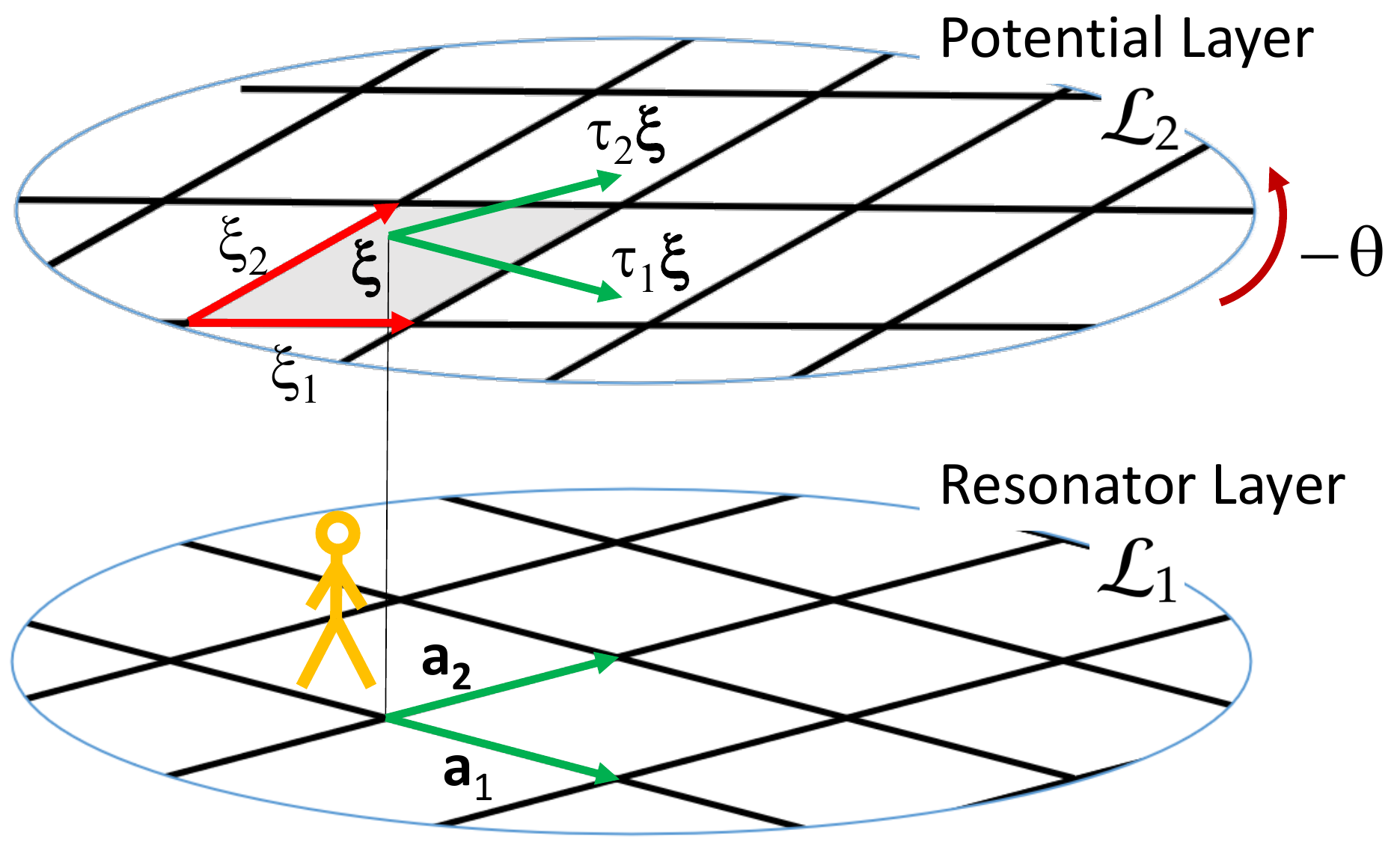}
\caption{\small {\bf Derivation of the phason space.} An observer can reproduce the entire bilayer from the coordinates $(\xi_1,\xi_2)$ of the origin of $\Ll_1$ relative to the shaded primitive cell of $\Ll_2$. If the observer moves to different resonators on the $\Ll_1$ lattice, this results in shifts over the folded $\RM^2/\Ll_2$ space generated by $\tau_1 ( \bm \xi)$ and $\tau_2 (\bm \xi)$.}
\label{Fig:DiscHull1}
\end{figure}

In the following, we compute this algebra explicitly and show that it is isomorphic to the non-commutative 4-torus. For this, we need first to parameterize the space of patterns. As in Fig.~\ref{Fig:DiscHull1}, it is useful to imagine an  observer sitting on top of a resonator. Looking around, one sees a certain pattern $\Pp$ and if the observer jumps to another resonator, say a hundred lattice units away, one will perceive a completely different pattern. The question is then, what is the minimum information the observer needs to reproduce the entire pattern, if we place the observer on top of an arbitrary resonator. Of course, the observer knows that one is dealing with a bilayer and that $\Ll_2$ is twisted by $\theta$ relative to $\Ll_1$. The answer is quite simple. The observer projects hers/his location onto the $\Ll_2$ plane and this projection $\bm \xi$ necessarily falls in  one primitive cell of $\Ll_2$. The observer sees the same pattern if this point lands on the opposite sides of the primitive cell, hence this primitive cell should be wrapped as a torus. In fact, the best strategy is to think that the entire second plane has been folded over one single primitive cell, e.g. the one shaded in Fig.~\ref{Fig:DiscHull1}. In other words, we work with the torus $\RM^2/\Ll_2$, which is parametrized as $\Xi = (\RM \, {\rm mod} \ a_1) \times (\RM \, {\rm mod} \ a_2)$, hence its points are $\bm \xi = (\xi_1,\xi_2)$, $\xi_i \in \RM \, {\rm mod} \ a_i$.

\begin{figure*}
\centering
\includegraphics[width=\linewidth]{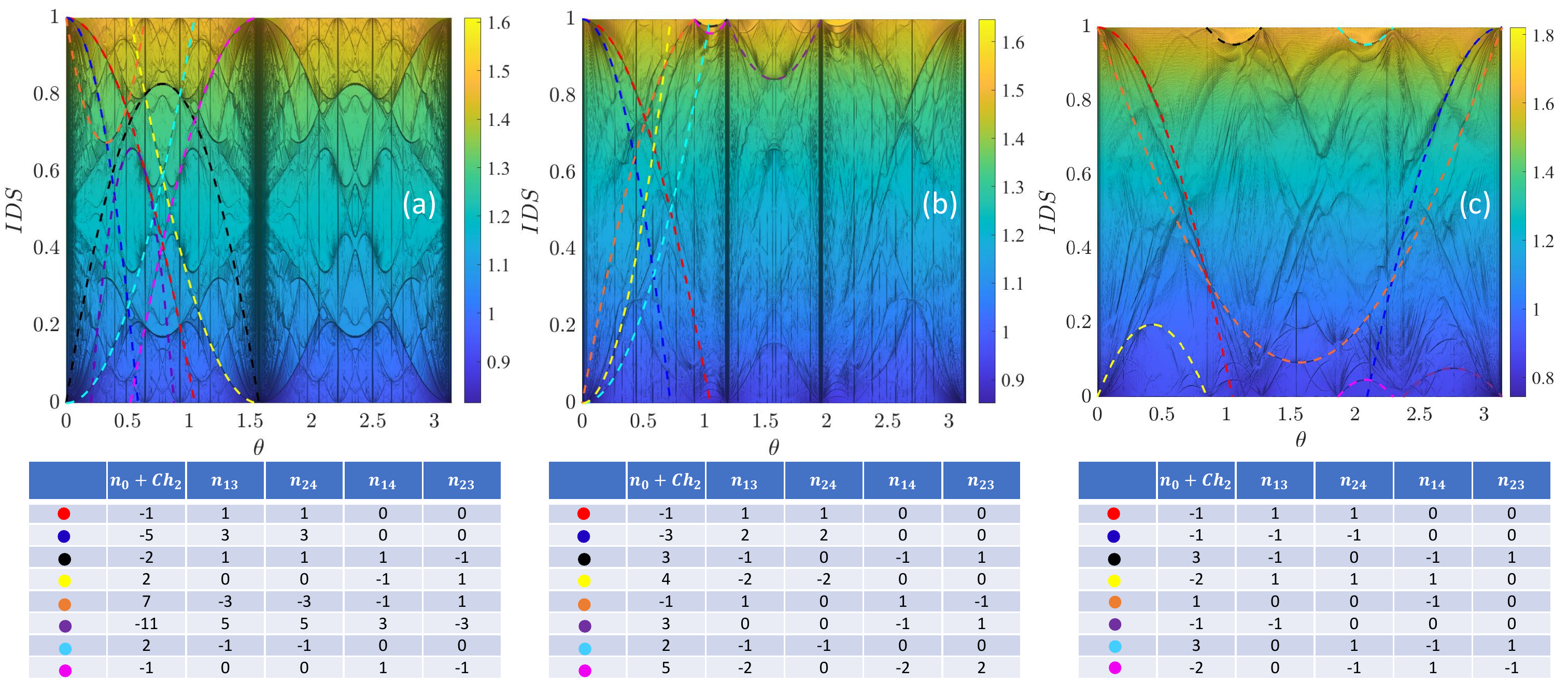}
\caption{\small {\bf Agreement between theory and simulation and computation of the topological invariants.} Integrated density of states (color maps), fittings by \eqref{Eq:IDSPredict} (dashed lines) and corresponding topological invariants (tables) for the bilayers (a) $a_1=a_2=1$ and $\beta=\pi/2$, (b) $a_1=a_2=1$ and $\beta=\pi/\sqrt{7}$ (b), and (c) $a_1=1/\sqrt[4]{2}$, $a_2=1$ and $\beta=\pi/\sqrt{7}$. The color-codings in panels (a), (b) and (c) are the same as in Figs.~\ref{Fig:SpecIDS1}(b), \ref{Fig:SpecIDS2}(b) and \ref{Fig:SpecIDS3}(b), respectively.}
\label{Fig:IDSfitting}
\end{figure*}

Now, the only information the observer needs in order to re-draw $\Pp$ is the position of its projection $\bm \xi$ in this primitive cell, hence $\bm \xi$ is the phason of the aperiodic pattern. Indeed, suppose that both layers have been erased. In this case, the observer will use $\bm a_i$ to redraw $\Ll_1$. Then, using the coordinates $(\xi_1,\xi_2)$ together with the given twist angle, the observer re-traces the primitive cell of $\Ll_2$ immediately above her/him and then tiles the plane by periodic translations of this primitive cell. An important question is if the observer explores the whole $\Xi$ or only a part of it, as she/he jumps from one resonator to another. Of course, this question is equivalent to asking if the dynamical system $\tau_a \Pp$ described above is topologically ergodic. 

From Fig.~\ref{Fig:DiscHull1}, we can see that these are just translations of the torus. In terms of coordinates $\xi_i$, the generators of these translations are given by
\begin{equation}\label{Eq:Translation}
\tau_{i} (\bm \xi) = \begin{pmatrix} (\xi_1 + a_1 A_{i1}) \, {\rm mod}\, a_1 \\ (\xi_2 + a_2 A_{i2}) \, {\rm mod} \, a_2\end{pmatrix}, \quad i=1,2,
\end{equation}
with:
\begin{equation}
A=
\begin{pmatrix} \frac{\sin(\beta-\theta)}{\sin(\beta)} & \frac{a_1}{a_2}\frac{\sin(\theta)}{\sin(\beta)} \\ 
-\frac{a_2}{a_1}\frac{\sin(\theta)}{\sin(\beta)} & \frac{\sin(\beta+\theta )}{\sin(\beta)}
\end{pmatrix} \quad ({\rm Det}\, A =1).
\end{equation}
As it is well known, if at least two $A_{ij}$'s are irrational numbers, then the orbit of one single point under repeated translations \eqref{Eq:Translation} fills the torus densely, hence the dynamical system $(\Xi,\tau)$ is topologically ergodic. One should not be surprised by the existence of this dynamical system because $(\Xi,\tau)$ is just the hull of $\Pp$, predicted to exist for any point pattern in \cite{Bellissard1986, Bellissard1995,KellendonkRMP95}.

We now have all the information to compute the algebra which generates the dynamical matrices. Since any continuous function over torus accepts a discrete Fourier decomposition, written slightly differently below,
\begin{align}
f(\bm \xi) = \sum_{\bm q \in \ZM^2} f_{\bm q} \ \Big (e^{\imath 2 \pi \xi_1/a_1}\Big)^{q_1} \Big(e^{\imath 2 \pi \xi_2/a_2}\Big)^{q_2},
\end{align}
the algebra of $T_f$ operators has two generators $T_j$ corresponding to the elementary functions:
\begin{equation}
f \mapsto u_j(\bm \xi) =e^{\imath 2\pi \xi_j/a_j}, \quad j=1,2.
\end{equation} 
Furthermore, since $(u_j \circ \tau_i)(\bm \xi) = e^{\imath 2\pi \, A_{ij}} u_j(\xi)$, the CR's \eqref{Eq:CommRel} become $
 S_iT_j= e^{-\imath 2\pi \, A_{ij}} T_j S_i$. As such, the algebra which contains all Galilean invariant dynamical matrices is generated by four unitary elements: $U_1=S_1$, $U_2=S_2$, $U_3=T_1$, $U_4=T_2$, with CR's $U_iU_j = e^{\imath 2 \pi \phi_{ij}} U_j U_i$, where
\begin{equation}
\Phi = \big [\phi_{ij} \big ] = {\small \begin{pmatrix} 
0 & 0 & -A_{11} & -A_{12} \\
0 & 0 & - A_{21} & -A_{22} \\
A_{11} & A_{21} & 0 & 0 \\
A_{12} & A_{22} & 0 & 0
\end{pmatrix}}
\end{equation}
Let us point out that, by considering two degrees of freedom per resonator, additional entries of the $\Phi$-matrix can be populated by $\pm \frac{1}{2}$ values, as explained in \cite{KrausPRL2013} for a spinful model.

{\it Predictions via K-Theory.} In K-Theory [75], the set of projections that can be (stably) deformed into a given projection is called the $K_0$-class of that projection. It is the complete topological invariant associated to that projection, in the sense any other topological invariant is already determined by its $K_0$-class. These classes of projections can be added and subtracted, hence they form an abelian group, the $K_0$-group of the algebra. Two homotopic projection $P$ and $P'$ are also similar: $P' = U^\ast P U$ for some unitary element from the same algebra. If $\Tt$ is a trace on the algebra, then automatically $\Tt(P) = \Tt(P')$ because we are allowed to make cyclic permutations inside a trace. This means that any trace is constant over the homotopy classes of projections. As such, traces are bona-fide  topological invariants.

\begin{figure*}
\centering
\includegraphics[width=\linewidth]{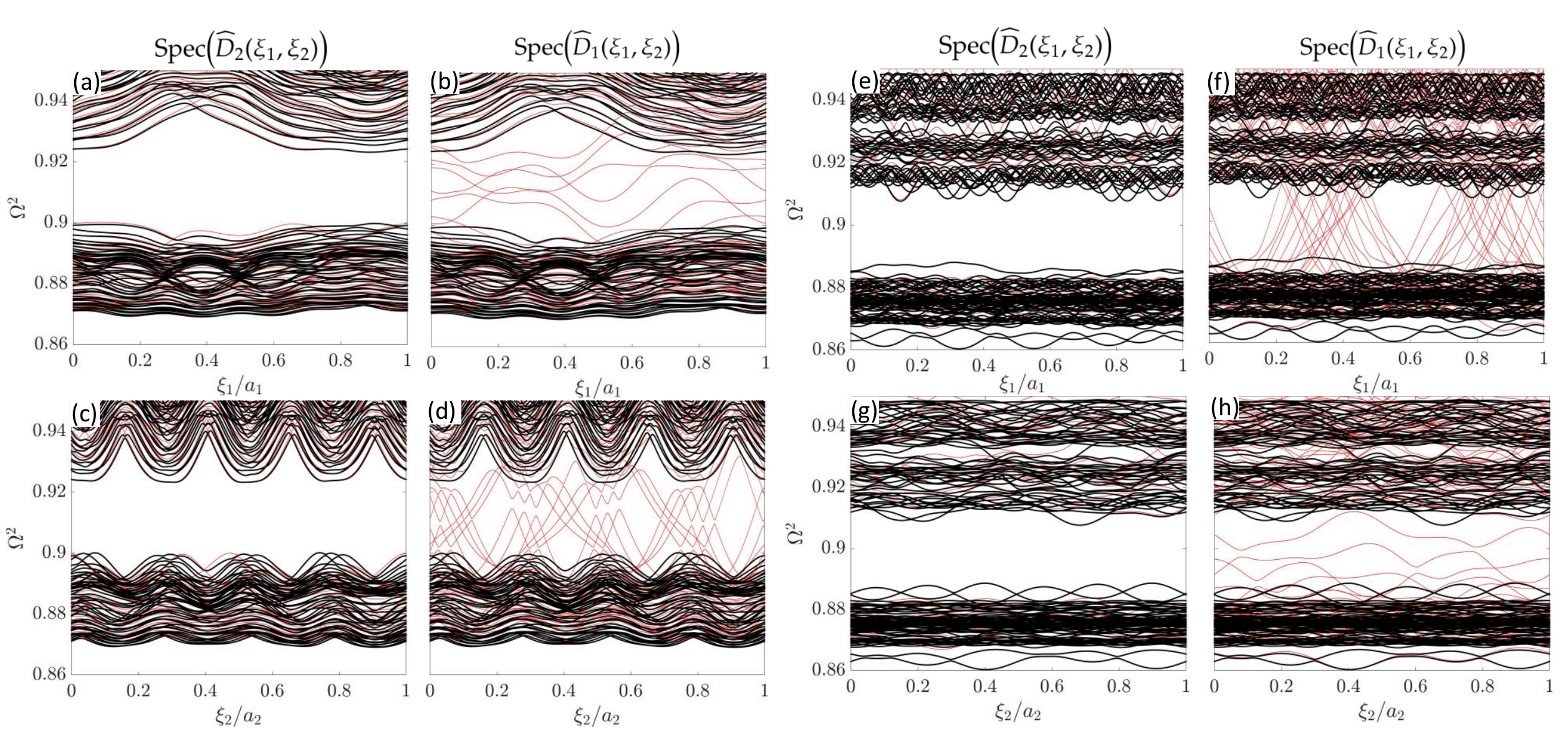}
\caption{\small {\bf Bulk-boundary correspondence 1.} Spectral flow w.r.t. $\xi_{1,2}$ of the dynamical matrix in the presence of a boundary for (a-d)  \textcolor{orange}{\large $\bullet$}-gap and (e-h) \textcolor{violet}{\large $\bullet$}-gap from Fig.~\ref{Fig:SpecIDS3}(b). The simulations were performed on a $21 \times 21$ lattice and $\theta = 1.55$ for \textcolor{orange}{\large $\bullet$}-gap, and on a $26 \times 26$ lattice and $\theta = 2.75$ for \textcolor{violet}{\large $\bullet$}-gap. The spectra computed with periodic boundary conditions (black curves) have been overlaid on top such that the boundary spectra (red curves) can be easily identified.}
\label{Fig:BB1}
\end{figure*}

In the case of non-commutative 4-torus, the $K_0$-group has $2^{d-1}$-generators $\{e_J\}$, conveniently labeled by a subset $J\subseteq \{1,2,3,4\}$ of directions with $|J| = {\rm even}$ \cite{FootnoteX}. Furthermore, for generic $\Phi$-matrices, the non-commutative 4-torus accepts a unique trace $\Tt$, which coincides with the trace per volume \cite{Bellissard1986}.  Any gap projection $P_G$ of a dynamical matrix defines a $K_0$-class and accepts a decomposition in terms of the generators $[P_G]_0 = \sum_{J} n_J \, [e_J]_0$. The integer numbers $n_J$ are called gap labels \cite{Bellissard1995} and, in general, they represent the complete set of independent topological invariants that can be associated to a gap projection. They are related but not necessarily equal to the Chern numbers (see below).

Since traces are linear maps,
\begin{equation}
\Tt[P_G]_0 = \sum_J n_J \, \Tt[e_J]_0.
\end{equation}
On the other hand,
\begin{equation}
\Tt(P_G)=\lim_{|\Ll_1| \rightarrow \infty} \frac{{\rm Tr}(P_G)}{|\Ll_1|} = \lim_{|\Ll_1| \rightarrow \infty} \frac{\{\# \ {\rm states} \ {\rm below} \ G\}}{|\Ll_1|},
\end{equation}
hence $\Tt(P_G) = {\rm IDS}(G)$. As such, if we can resolve the values of the trace on the generators of the $K_0$-group, we can make a prediction about the allowed values of ${\rm IDS}$. For the non-commutative torus, this extremely useful piece of information was supplied in \cite{Elliott1984}, and we have
\begin{equation}
{\rm IDS}(G) = \Tt(P_G) = \sum\limits_{J \subset \{1,2,3,4\}}^{|J|={\rm even}} n_J \, {\rm Pf}(\Phi_J),
\end{equation}
 where $\Phi_J$ is the matrix $\Phi$ restricted to indices $J$ and ${\rm Pf}$ is the pfaffian of the resulting anti-symmetric matrix.  In our case, this gives the prediction \cite{Footnote2}
\begin{align}\label{Eq:IDSPredict}
{\rm IDS(G)} =  n_{\emptyset} & + n_{\{1,3\}} A_{11}+n_{\{1,4\}} A_{12} + n_{\{2,3\}} A_{21} \\ \nonumber
&  + n_{\{2,4\}} A_{22}   + n_{\{1,2,3,4\}} \, {\rm Det}(A).
\end{align}
When there are no linear relations with integer coefficients between $A_{ij}$-s, we can compute all topological invariants supplied by $n_{\{i,j\}}$-s by fitting \eqref{Eq:IDSPredict} to the numerically obtained ${\rm IDS}$ curves in Figs.~\ref{Fig:SpecIDS1}(c), \ref{Fig:SpecIDS2}(c) and \ref{Fig:SpecIDS3}(c). Unfortunately, ${\rm Det}(A) = 1$, hence we can only determine the sum $n_\emptyset + n_{\{1,2,3,4\}}$ via this procedure. 

The values of the Chern numbers on the $K_0$-generators were computed in \cite{ProdanSpringer2016}[p.~141]:
\begin{equation}\label{Eq:ChernValues}
{\rm Ch}_{J'} [e_J]_0 = \left \{ 
\begin{array}{l}
0 \ {\rm if} \ J'\nsubseteq J  , \\
1 \ {\rm if} \ J' = J , \\
{\rm Pf}(\Phi_{J\setminus J'} \ {\rm if} \ J' \subset J,
\end{array}
\right .  \quad J, J' \subset \{1,2,3,4\}.
\end{equation}
Since the Chern numbers are also linear maps, their values on the gap projection $[P_G]_0 = \sum_{J} n_J \, [e_J]_0$ can be straightforwardly computed from \eqref{Eq:ChernValues}:
\begin{equation}
{\rm Ch}_{J'}[P_G]_0 = n_{J'} + \sum_{J' \subsetneq J} n_J \, {\rm Pf}(\Phi_{J\setminus J'}).
\end{equation} 
As one can see, the top Chern number corresponding to $J'=\{1,2,3,4\}$, also known as the second Chern number and denoted by ${\rm Ch}_2$, is always an integer, but the lower Chern numbers may not be. We will use the above relations in our discussion of the bulk-boundary correspondence.

\subsection{Theory meets numerics} \label{Sec:ThVsNum}

For the bilayer analyzed in Fig.~\ref{Fig:SpecIDS1},  there are linear dependencies and the prediction from Eq.~\ref{Eq:IDSPredict} reduces to
\begin{align}\label{Eq:PredictedIDS1}
{\rm IDS}(G)=(n_{\emptyset}+{\rm Ch}_2(P_G))&+(n_{\{1,3\}}+n_{\{2,4\}})\cos\theta \\ \nonumber
&+(n_{\{1,4\}}-n_{\{2,3\}})\sin\theta.
\end{align}
However, due to the symmetry under $\pi/2$ rotations, $n_{\{1,3\}}=n_{\{2,4\}}$ and $n_{\{1,4\}}=-n_{\{2,3\}}$, which follows directly from the expressions of the first Chern numbers. As such, \eqref{Eq:PredictedIDS1} can be used to determine all topological invariants supplied by $n_{\{i,j\}}$. We found that Eq.~\eqref{Eq:PredictedIDS1} fits perfectly all IDS curves seen in Fig~\ref{Fig:SpecIDS1}(c). Fittings of the IDS curves inside the eight large gaps identified in Fig.~\ref{Fig:SpecIDS1}(b) are reported in Fig.~\ref{Fig:IDSfitting}(a), together with the topological invariants extracted from the fittings. Let us remark that, by using the symmetries of the spectral butterfly, we can automatically fit many more IDS curves, 56 to be more precise.

\begin{figure}
\centering
\includegraphics[width=\linewidth]{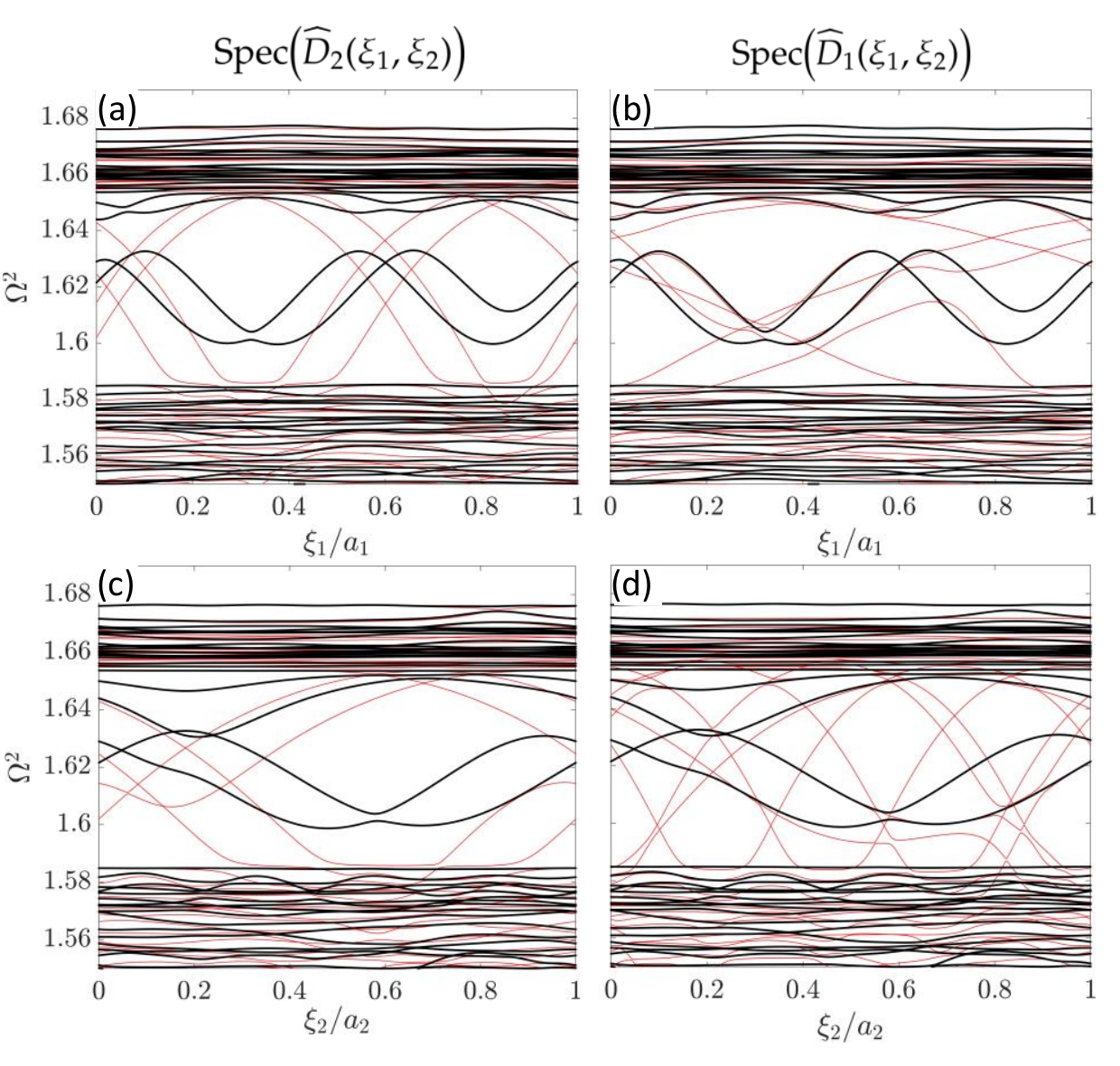}
\caption{\small {\bf Bulk-boundary correspondence 2.} Same as Fig.~\ref{Fig:BB1}(a-d) for \textcolor{red}{\large $\bullet$}-gap from Fig.~\ref{Fig:SpecIDS3}(b).}
\label{Fig:BB2}
\end{figure}

For the bilayer analyzed in Fig.~\ref{Fig:SpecIDS2}, one additional linearly independent term is present in the IDS expression:
\begin{align}\label{Eq:PredictedIDS2}
{\rm IDS}(G)=(n_{\emptyset}&+{\rm Ch}_2(P_G))+n_{\{1,3\}}\frac{\sin(\beta-\theta)}{\sin\beta} \\ \nonumber
&+(n_{\{1,4\}}-n_{\{2,3\}})\frac{\sin\theta}{\sin\beta} + n_{\{2,4\}}\frac{\sin(\beta+\theta)}{\sin\beta}.
\end{align}
Form symmetry considerations, we found again that $n_{\{1,4\}} = -n_{\{2,3\}}$.  Again, we have verified that Eq.~\eqref{Eq:PredictedIDS2} perfectly fits all IDS curves seen in Fig~\ref{Fig:SpecIDS2}(c). Fittings of the IDS curves inside the eight large gaps identified in Fig.~\ref{Fig:SpecIDS2}(b) are reported in Fig.~\ref{Fig:IDSfitting}(b), together with the topological invariants extracted from the fitting. The symmetry of the spectral butterfly can be used to automatically fit eight additional IDS curves in the right side of the IDS plot.

Finally, for the bilayer analyzed in Fig.~\ref{Fig:SpecIDS3}, we have five linearly independent terms present in the IDS expression:
\begin{align}\label{Eq:PredictedIDS3}
{\rm IDS}(G)=&(n_{\emptyset}+{\rm Ch}_2(P_G))+n_{\{1,3\}}\frac{\sin(\beta-\theta)}{\sin\beta} \\ \nonumber
&+ n_{\{2,4\}}\frac{\sin(\beta+\theta)}{\sin\beta}+ n_{\{1,4\}}\frac{a_1}{a_2}\frac{\sin\theta}{\sin\beta} - n_{\{2,3\}}\frac{a_2}{a_1}\frac{\sin\theta}{\sin\beta}.
\end{align}
In this case there is no point symmetry left and the four topological numbers $n_{\{i,j\}}$ are all independent. We found again that Eq.~\eqref{Eq:PredictedIDS2} fits perfectly all IDS curves seen in Fig~\ref{Fig:SpecIDS3}(c). Fittings of the IDS curves inside the eight large gaps identified in Fig.~\ref{Fig:SpecIDS3}(b) are reported in Fig.~\ref{Fig:IDSfitting}(c), together with the topological invariants extracted from the fitting.

Let us point out that every single gap among the 24 gaps analyzed in Fig.~\ref{Fig:IDSfitting} displays a non-zero $n_{\{i,j\}}$ but we have not yet able to demonstrate the existence on nontrivial top invariants $n_{\{1,2,3,4\}}$. For this, we turn to the bulk-boundary correspondence for the twisted bilayers. 

\subsection{Bulk-boundary correspondence and existence of $2^{\rm nd}$-Chern states}

Physical boundaries are created by restricting either one of $n_k$ coefficients of $\bm r_{\bm n}$ to non-negative values. If $n_k \geq 0$, then the boundary cuts the $k$-th direction and we will call it a $k$-boundary. We denote the resulting dynamical matrix by $\widehat D_k(\bm \xi)$. The bulk-boundary for class A in higher dimensions states \cite{ProdanSpringer2016} that the surface states admit topological invariants in the form of odd Chern numbers and that there is a precise relation between all bulk and surface invariants. In particular, for our lower Chern numbers \cite{ProdanSpringer2016}[p.~175],
\begin{equation}
{\rm Ch}_{\{k,i\}}(P_G)= \left . \frac{N_{ki}}{\sqrt{|\Ll_1|}}\right |_{|\Ll_1| \rightarrow \infty}, \quad k\in \{1,2\}, \ i\in \{3,4\},
\end{equation}
where $N_{ki}$ is the net number \cite{Footnote3} of eigenvalues of $\widehat D_k(\xi_1,\xi_2)$ that cross an arbitrary reference line inside the bulk gap when $\xi_i$ is varied from $0$ to $a_i$ while holding the other $\xi$ fixed. Using \eqref{Eq:ChernValues}, we can write the bulk-boundary principle explicitly,
\begin{equation}\label{Eq:BBExplicit}
\left . \frac{N_{ki}}{\sqrt{|\Ll_1|}}\right |_{|\Ll_1| \rightarrow \infty}= n_{\{1,2,3,4\}}\, {\rm Pf}\big(\Phi_{\{1,2,3,4\}\setminus\{k,i\}}\big) +n_{\{k,i\}}.
\end{equation}

Numerically, we generate a $k$-boundary by using open boundary conditions in that physical direction and periodic boundary condition in the remaining direction. As always, we will create a pair of boundaries, hence the numerically computed edge modes will always come in pairs. In Fig.~\ref{Fig:BB1}(a-d), which was simulated on a $21 \times 21$ lattice, we analyze the bulk-boundary correspondence for the \textcolor{orange}{\large $\bullet$}-gap from Fig.~\ref{Fig:SpecIDS3}(b). As one can see, $\widehat D_1(\bm \xi)$ displays 21 positively sloped chiral bands when $\xi_2$ is varied \cite{Footnote4}, and no chiral bands are present for the other three cases. This is consistent with $n_{\{1,4\}}=1$ and trivial values for the other invariants. Similarly, in Fig.~\ref{Fig:BB1}(e-f), which was simulated on a $26 \times 26$ lattice, we analyze the bulk-boundary correspondence for the \textcolor{violet}{\large $\bullet$}-gap from Fig.~\ref{Fig:SpecIDS3}(b). In this case, $\widehat D_1(\bm \xi)$ displays 27 positively sloped chiral bands when $\xi_1$ is varied and no chiral bands are present for the other three cases. This is consistent with $n_{\{1,3\}}=1$ and trivial values for the other invariants. These numerical findings confirm the predicted bulk-boundary correspondences based on \eqref{Eq:BBExplicit} and the data from Fig.~\ref{Fig:IDSfitting} and, furthermore, they enable us to actually conclude that $n_{\{1,2,3,4\}}=0$ for these two particular gaps.

Additional boundary spectra are reported in Fig.~\ref{Fig:BB2}, which were simulated on a $23 \times 23$ lattice and $\theta=0.2$. They correspond to the \textcolor{red}{\large $\bullet$}-gap in Fig.~\ref{Fig:SpecIDS3}(b). From the data in Figs.~\ref{Fig:BB2} and \ref{Fig:IDSfitting}, and by assuming $n_{\{1,2,3,4\}}=-1$, we have:
\begin{align}
&\frac{N_{23}=4}{23}=0.17, \ |{\rm Ch}_{\{2,3\}}|=|-{\rm Pf}(\Phi_{\{1,4\}})+n_{\{2,3\}}|=0.18; \nonumber \\ \nonumber
&\frac{N_{13}=2}{23}=0.09, \ |{\rm Ch}_{\{1,3\}}|=|-{\rm Pf}(\Phi_{\{24\}})+n_{\{1,3\}}|=0.06; \\ \nonumber
&\frac{N_{24}=2}{23}=0.09, \ |{\rm Ch}_{\{2,4\}}|=|-{\rm Pf}(\Phi_{\{1,3\}})+n_{\{1,3\}}|=0.1; \\ 
&\frac{N_{14}=6}{23}=0.26, \ |{\rm Ch}_{\{1,4\}}|=|-{\rm Pf}(\Phi_{\{2,3\}}+n_{\{1,4\}}|=0.25.
\end{align}
As one can see, the predicted bulk-boundary correspondence \eqref{Eq:BBExplicit} holds with a remarkable precision given the relatively small size of the lattice \cite{Footnote5}. Similar agreements hold true for other spectral gaps from Fig.~\ref{Fig:SpecIDS3}(b) and, for example, for the \textcolor{cyan}{\large $\bullet$}-gap we found $n_{\{1,2,3,4\}}=1$. As such, twistronics is capable of generating gaps with $2^{\rm nd}$-Chern numbers.
 
\section{Discussion}  

We have demonstrated that twistronics can be a simple yet extremely effective way to produce topological gaps and topological boundary modes. Indeed, twisted bilayers have a  ``hidden'' degree of freedom, the phason $\bm \xi$, which leaves on a torus and can be controlled by simple relative shifts of the layers. For generic twist angles, these shifts do not affect the bulk spectrum, hence the bulk gaps, but they generate dispersive chiral boundary modes in the presence of a boundary. The count of these modes agrees with a precise topological bulk-boundary principle.

To our knowledge, this is the first time when the algebra of dynamical matrices for a Moir\'e pattern has been explicitly computed. With that result at hand, the K-theoretic machinary invented by Bellissard \cite{Bellissard1986} enabled us to classify all topological phases from class A supported by these twisted bilayers and to produce a high-throughput of topological gap labels. Let us mention that the topological invariants can be computed directly using the algorithms developed in \cite{ProdanSpringer2017}. However, those algorithms require a substantial computational effort and, as such, the technique based on the IDS fitting is an important outcome of our work. Using the newer results from \cite{ProdanSpringer2016}, we were able to also make precise predictions about the bulk-boundary correspondence for these Moir\'e patterns. To our knowledge, it is the first time when a bulk-boundary correspondence is observed for non-integer invariants. 

Our analysis generalizes to the cases where there are more degrees of freedom per primitive cell, such as the honeycomb lattice, or when the coupling constants are modulated not by one but by multiple twisted lattices. Indeed, assume that a lattice $\Ll_3$ is added on top of $\Ll_2$ in Fig.~\ref{Fig:DiscHull1}. By following similar arguments, it is easy to see that reproducing the whole patterns requires the knowledge of the projection of the resonator where the observer sits on both $\RM^2/\Ll_2$ and $\RM^2/\Ll_3$ tori. As such, the phason space is a 4-torus and the dynamical system $\tau$ can be computed by similar methods. It follows that the algebra which generates the dynamical matrices is the non-commutative 6-torus, which hosts topological phases with $3^{\rm rd}$-Chern numbers \cite{ProdanSpringer2016}.

In conclusion, we proved that the twisted layered systems, both classical and quantum, can be resourceful virtual laboratories for exploring completely new physics and topological states. For metamaterials research, our findings open new venues for engineering topological gaps and robust boundary modes without any need for fine-tuning or active components. 

{\bf Data Availability.} The data and the numerical codes are available from the authors upon request.

\acknowledgments{M. Rosa and M. Ruzzene gratefully acknowledge the support from the National Science Foundation (NSF) through the EFRI1741685 grant and from the Army Research Office through grant W911NF-18-1-0036. E. Prodan acknowledges support from the W. M. Keck Foundation and from National Science Foundation through the grant DMR-1823800.}


\begin{thebibliography}{}
\bibitem{FootnoteX} The empty set $\emptyset$ is also a valid choice for $J$.
\bibitem{Footnote2} Technically, the Pfaffians should be positive but relaxing the signs makes the fitting easier. For bulk-boundary correspondence, however, one need to re-assess the signs of the terms.
\bibitem{Footnote3} The eigenvalues which cross the reference point from below/above are counted with $\pm$ signs, respectively.
\bibitem{Footnote4} The small gaps are due to the hybridization of the boundary modes from opposite sides and should be neglected.
\bibitem{Footnote5} We purposely kept the lattice size small to be able to count the edge bands.
\end{thebibliography}
\end{document}